\documentclass[paper]{ws-ijmpcs}

\usepackage{mciteplus}
\graphicspath{figs}

\newcommand{\cp}{{$\mathcal{CP}$}}
\newcommand{\GeV}{{\;\mathrm{GeV}}}
\newcommand{\mhmax}{{$M_h^{\mathrm{max}}$}}
\newcommand{\mhmod}{{$M_h^{\mathrm{mod}}$}}
\newcommand{\lowmh}{{\emph{low-$M_H$}}}
\newcommand{\FH}{{\tt FeynHiggs}}
\newcommand{\HB}{{\tt HiggsBounds}}
\newcommand{\HS}{{\tt HiggsSignals}}

\begin{document}

\markboth{O.~St{\aa}l}{Higgs Boson Scenarios Beyond the Standard Model}

\catchline{}{}{}{}{}
%
%%%%%%%%%%%%%%%%%%%%%%%%%%%%%%%%%%%%%%%%%%%%%%%%%%%%%%%%%%%%%%%%%%%%

\title{PROSPECTS FOR HIGGS BOSON SCENARIOS\\ BEYOND THE STANDARD MODEL}

\author{OSCAR ST{\AA}L}

\address{The Oskar Klein Centre, Department of Physics\\
Stockholm University, SE-106 91, Stockholm, Sweden\\
oscar.stal@fysik.su.se}

\maketitle

%\begin{history}
%\received{26 Feb 2014}
%\revised{Day Month Year}
%\end{history}

\begin{abstract}
The new particle recently discovered at the Large Hadron Collider has properties compatible with those expected for the Standard Model (SM) Higgs boson. However, this does not exclude the possibility that the discovered state is of non-standard origin, as part of an elementary Higgs sector in an extended model, or not at all a fundamental Higgs scalar. We review briefly the motivations for Higgs boson scenarios beyond the SM, discuss the phenomenology of several examples, and summarize the prospects and methods for studying interesting models with non-standard Higgs sectors using current and future data.
\end{abstract}

%\ccode{PACS numbers:}

\section{Introduction}	
After the 2012 discovery of a new particle in the $\gamma\gamma$ and $ZZ$ final states by ATLAS\cite{ATLASDiscovery} and CMSÊ\cite{CMSDiscovery}, experimental Higgs physics has entered a new era. Using already up to the full dataset from the first LHC run, corresponding to an integrated luminosity (per experiment) of $\sim 5$ fb$^{-1}$ at $\sqrt{s}=7$~TeV and $\sim 25$ fb$^{-1}$ at $\sqrt{s}=8$~TeV, measurements of key properties for this new state have been reported by both collaborations\cite{ATLAS-CONF-2013-034,CMS-PAS-HIG-13-005}. To set the stage for the ensuing discussion on Higgs sectors beyond the SM, we summarize the key experimental findings here. The new particle has a measured mass of\cite{CMS-PAS-HIG-13-005,ATLAS-CONF-2013-014}
\begin{equation*}
\begin{aligned}
M_H&=125.5\pm 0.2(\mathrm{stat.})^{+0.5}_{-0.6}(\mathrm{syst.})\GeV\quad \mathrm{(ATLAS)}, \\
M_H&=125.7\pm 0.3(\mathrm{stat.})\pm 0.3(\mathrm{syst.})\GeV\quad \mathrm{(CMS)}. \\
\end{aligned}
\end{equation*}
Although the SM Higgs mass is in principle a free parameter, the measured value is in good agreement with earlier predictions performed using electroweak precision data. With the accurate Higgs mass determination, the fate of the SM vacuum has been analyzed to great precision\cite{Degrassi:2012ry}, with the outcome that the SM can be extrapolated to very high scales ($Q \sim 10^{11}~\GeV$); the SM vacuum is likely metastable with a lifetime exceeding the age of the universe.

\begin{figure}
\centering
\includegraphics[width=0.42\columnwidth]{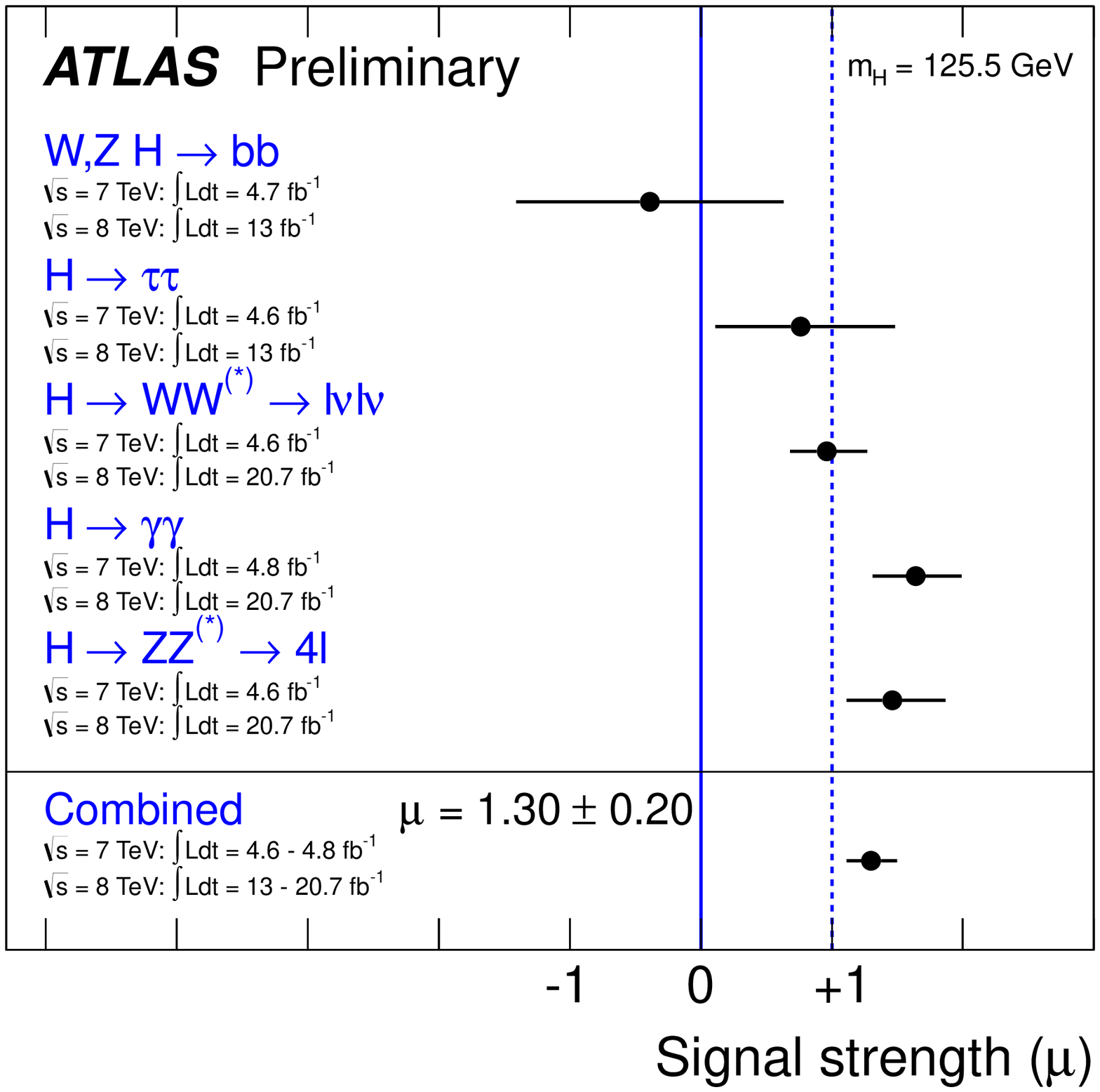}
\includegraphics[width=0.43\columnwidth]{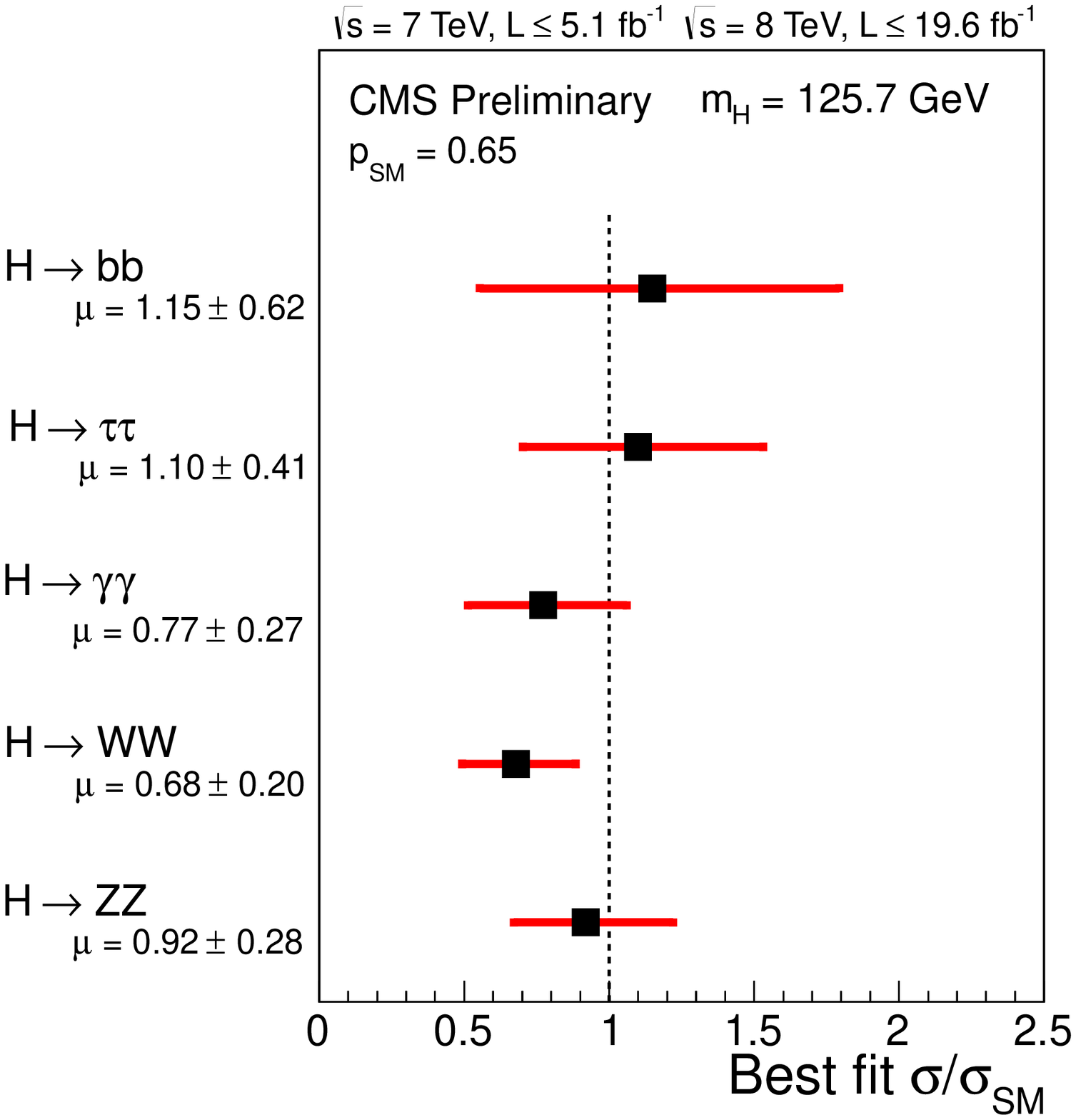}
\caption{Higgs signal strengths, relative to the SM value $\mu_i=1$, measured in different final states by ATLAS\protect\cite{ATLAS-CONF-2013-034} (left) and CMS\protect\cite{CMS-PAS-HIG-13-005} (right).}
\label{fig:SMcouplings}
\end{figure}

In the SM, the Higgs signal rates ($\sigma\times\mathrm{BR}$) into various channels are completely fixed by the Higgs mass. Precise predictions in various channels can then be combined with data to measurements of individual signal strenghts, $\mu_i$, normalized such that the SM corresponds to $\mu_i=1$. The current status of the Higgs signal strengths measurements is shown in Fig.~\ref{fig:SMcouplings}.
As can be seen from this figure, there is currently no statistically significant deviation from $\mu_i=1$ observed in the data. 

To establish that the observed particle is indeed a fundamental scalar, as predicted by the Higgs mechanism, it is also necessary to determine its spin-parity quantum numbers, $J^P$ (the SM Higgs has $J^P=0^+$). First attempts at this have rejected minimal alternative hypotheses like (pure) $0^-$ or $2^+$ states at the $95\%$ confidence level (CL)\cite{ATLAS-CONF-2013-040,Chatrchyan:2012jja}, which adds further credibility to the conclusion that the observed state is indeed a $0^+$ scalar, exactly as predicted by the SM. However, it should be remembered that this is based on much less conclusive evidence than the discovery itself and that there is still room for, e.g., a sizeable \cp-odd admixture. 

That said, we now turn to the question what alternative theories may be behind this discovery? After all, the SM is known to have  several shortcomings, such as the gauge hierarchy problem, the absence of a suitable dark matter candidate and, ultimately, the problem of how to incorporate gravity into the theory. Even if the Higgs data currently tell us that the discovered scalar boson fits nicely the minimal Higgs structure of the SM, it is well-known that several extensions incorporate the SM Higgs sector in a particular (decoupling) limit, and that the observations therefore make perfect sense also in these theories. There are two principal ways in which a non-minimal Higgs sector would manifest itself experimentally: either directly through the discovery of additional scalar states, which would prove immediately that the Higgs sector is non-minimal, or through precise measurements of the Higgs properties, that could eventually reveal deviations from the SM predictions for the discovered $125\GeV$ state.
Below we shall describe a few examples of physics beyond the SM where both strategies could be relevant.

\section{Supersymmetry}
Low-energy supersymmetry is widely recognized as an attractive extension of the SM, since the symmetry naturally protects the scalar Higgs mass against large quantum corrections. Other advantages of the minimal supersymmetric Standard Model (MSSM) is that it can offer unification of the gauge couplings and radiative generation of the electroweak scale. On the more phenomenological side, the MSSM with $\mathcal{R}$-parity conservation can provide a stable candidate particle for the cold dark matter in the lightest neutralino. Unfortunately, LHC searches for supersymmetric particles have so far turned up nothing\cite{ATLASSusyWiki}, which starts to put the naturalness arguments under question. Current limits, which imply for example that the gluino mass $m_{\tilde{g}}\gtrsim 1.2$~TeV, introduce already a certain degree of unavoidable fine-tuning\cite{Brust:2011tb,*Papucci:2011wy}. This adds to the potential fine-tuning arising from the Higgs mass measurement, which we shall discuss in more detail below.

\subsection{A $\mathit{ 125\GeV}$ Higgs boson in minimal SUSY}
The MSSM, which in many cases also serve as a template for weakly-coupled extended Higgs sectors, contains two complex Higgs doublets. Following electroweak symmetry breaking, there are five physical Higgs bosons in the spectrum. When \cp\ is conserved, they are classified as two \cp-even scalars, $h$ and $H$ ($M_h<M_H$ and the two states mix with an angle $\alpha$), one \cp-odd scalar, $A$, and a charged Higgs pair, $H^\pm$. At tree level, the Higgs sector can be specified using only two parameters, conveniently chosen as either $M_A$ or $M_{H^\pm}$ and $\tan\beta$, the ratio of vacuum expectation values of the two doublets. The remaining Higgs masses and the mixing angle $\alpha$ become predictions of the theory; in particular one finds that $M_h^{\mathrm{tree}} \leq M_Z$, with equality in the decoupling limit ($M_A\gg M_Z$, $\tan\beta\gg 1$) where the couplings of $h$ approach their SM values.  

Beyond leading order, large radiative corrections to the Higgs masses arise, in particular from the coupling to the top quark. The dominant corrections to the lightest Higgs mass at one loop can be written as
\begin{equation}
M_h^2 = M_Z^2+\frac{3m_t^4}{2\pi^2 v^2}\left[\log\frac{M_S^2}{m_t^2}+\frac{X_t^2}{M_S^2}\left(1-\frac{12X_t^2}{M_S^2}\right)\right],
\label{eq:mh1loop}
\end{equation}
where $M_S=\sqrt{m_{\tilde{t}_1}m_{\tilde{t}_2}}$ and $X_t=A_t-\mu\cot\beta$ is the  mixing parameter appearing in the off-diagonal entries of the scalar top (stop) mass matrix. According to Eq.~\ref{eq:mh1loop}, the lightest Higgs mass can be made heavier than $M_Z$ (and even reach $M_h\sim 125\GeV$), either by increasing the average stop mass, $M_S$, or by having a large mixing in the stop sector; the maximum attained for $X_t=\pm\sqrt{6}\,M_S$. In theories like the MSSM, where the Higgs masses are predictions, constraints on the model parameters can be derived from the Higgs mass value measured at the LHC\cite{Hall:2011aa,Heinemeyer:2011aa,Baer:2011ab,Arbey:2011ab,*Draper:2011aa}. The resulting constraints are particularly severe for high-scale models with mediation mechanisms that do not generate large $A$-terms\cite{Arbey:2011ab,*Draper:2011aa,Brummer:2012ns}.

In a bottom-up approach, the low-energy MSSM parameters can be treated as input directly without reference to a ``higher'' model with e.g.~grand unification. This fixes the radiative corrections to Higgs masses and mixing, and allows higher-order SQCD corrections to Higgs production and decay to be calculated. With a complete spectrum, all aspects of the model phenomenology can be studied. This is the basis for MSSM Higgs sector benchmark scenarios\cite{Carena:1999xa}, which were successfully used at LEP\cite{Schael:2006cr} and the Tevatron. With the Higgs discovery, many of these original scenarios have become obsolete, and updated scenarios for LHC Higgs phenomenology were recently proposed\cite{Carena:2013qia}. We are going to use some of these scenarios here to discuss the prospects for current and future MSSM Higgs searches.

\begin{figure}
\centering
\includegraphics[width=0.46\columnwidth]{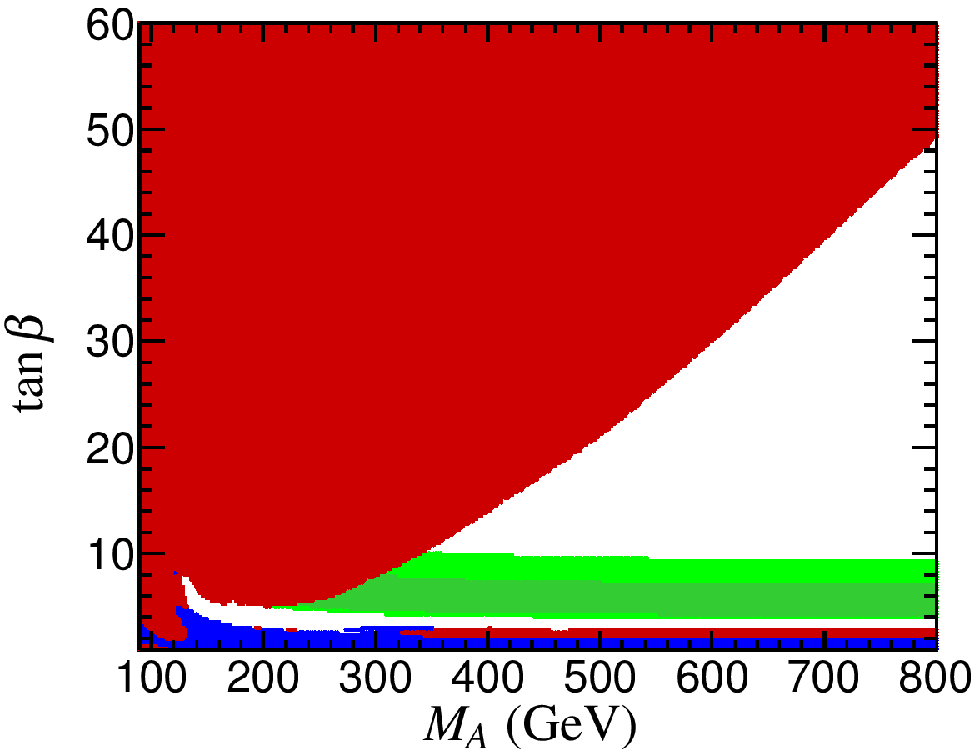}
\includegraphics[width=0.46\columnwidth]{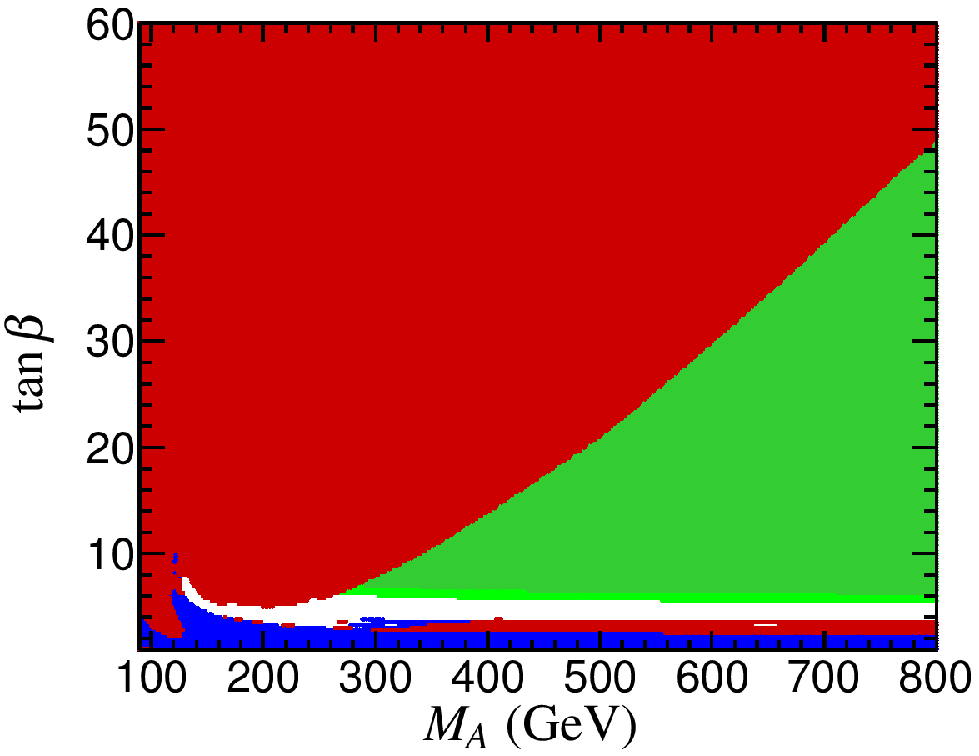}
\caption{Exclusion limits from direct Higgs searches at LEP (blue) and the LHC (red) in the MSSM benchmark scenarios \mhmax\ (left) and \mhmod\ (right). The dark (bright) green regions are compatible with the direct limits and has the lightest MSSM Higgs boson in the mass range $M_h=125.5\pm 2(3)\GeV$. The results have been produced using \FH\ and \HB.}
\label{fig:HBlimits}
\end{figure}

Numerical tools are available to calculate the MSSM Higgs spectrum to the two-loop level, which is necessary for precision phenomenology. As an example, we use numerical predictions from \FH\cite{Heinemeyer:1998yj,*Heinemeyer:1998np,*Degrassi:2002fi,*Frank:2006yh} (version 2.9.4) for two benchmark scenarios: \mhmax, which maximizes $M_h$ for given values of $M_S$, $M_A$ and $\tan\beta$, and \mhmod, in which the stop mixing is selected to maximize the region in $\tan\beta$ that is compatible with the observed Higgs mass value. The current experimental status for these scenarios is shown in Fig.~\ref{fig:HBlimits}. In this figure, direct search limits from LEP, Tevatron and the LHC are evaluated using \HB\cite{Bechtle:2008jh,*Bechtle:2011sb,*Bechtle:2013wla} (version 4). \HB\ is a convenient program to take into account limits from direct Higgs searches in arbitrary models. Comparing the exclusion limits for the two scenarios in Fig.~\ref{fig:HBlimits}, several features are common. The bulk exclusion at high $\tan\beta$ comes from combined MSSM $h/H/A\to \tau\tau$ searches\cite{CMSHig12050}, a channel which receives a strong $\tan\beta$ enhancement. The same parameter region at low $M_A$, high $\tan\beta$ is also the one most strongly constrained by indirect measurements, such as e.g.~$B_s\to \mu^+\mu^-$\cite{Aaij:2013aka}, and this region therefore appears strongly disfavoured in the MSSM. This conclusion is corroborated by global analyses of the (low-energy) ``pMSSM'' parameter space taking into account the measured Higgs rates\cite{Cao:2012yn,*Bechtle:2012jw,*Arbey:2012bp}, which now favour the region $M_A\gtrsim 300\GeV$.\cite{Djouadi:2013lra} Finally, we note from Fig.~\ref{fig:HBlimits} that for SUSY-breaking masses $M_S$ around the TeV-scale (here: $M_S=1\;\mathrm{TeV}$), the low $\tan\beta$ region is not accessible for any value of $M_A$, since $M_h$ becomes too low (even below the LEP limit).

The global MSSM analyses are useful to identify interesting phenomenology which can still be viable in the allowed parameter space regions. For example, while the jury is still out on the possibility of an enhancement in the two-photon signal in the data (as weakly indicated by ATLAS, but not supported by CMS, see Fig.~\ref{fig:SMcouplings}), it is useful to assess under which circumstances such an enhancement can be reproduced. In the MSSM, $h\to \gamma\gamma$ can be enhanced primarily in two ways: either by suppressing the main decay mode, $h\to b\bar{b}$, through mixing between the two Higgs doublets, or by direct contributions to the partial width $\Gamma(h\to \gamma\gamma)$ from new particles. The largest possible new physics contribution comes from the scalar taus (staus)\cite{Carena:2011aa,*Carena:2012gp}, which unlike a contribution from e.g.~stops would not suppress Higgs production through gluon fusion since the staus do not carry colour charge. It is found that an enhancement of $R_{\gamma\gamma}=(\sigma\times \mathrm{BR})/\mathrm{SM}\gtrsim 1.5$ is possible if $m_{\tilde{\tau}_1}\sim 100$~GeV, close to the current limit. 

Due to the high degree of compatibility between measurements and the SM, the current Higgs data favours the decoupling limit ($M_A \gg M_Z$, $\tan\beta\gg 1$), in which the couplings of $h$ become SM-like. Nevertheless, MSSM Higgs phenomenology could be strikingly different from that of the SM, in particular due to the presence of additional, heavier, Higgs states, which can still be as light as $\sim 300$~GeV. This may also lead to interesting possibilities for interactions with the light Higgs through, for example, the cascade decays $H\to hh$ and $A\to hZ$ which could be significant.  The MSSM Higgs bosons could also be produced in SUSY cascades, such as e.g.~$\tilde{\chi}_i^0 \to \tilde{\chi}_j^0 h$. Finally, if there are light EW SUSY states, the heavy MSSM Higgs bosons may predominantly decay into light -ino pairs, $H/A \to \tilde{\chi}_i^0 \tilde{\chi}_j^0$, leading to quite unusual signatures. These options should be considered for future search strategies.

\subsection{Heavy Higgs interpretation}
\begin{figure}
\centering
\includegraphics[width=0.48\columnwidth]{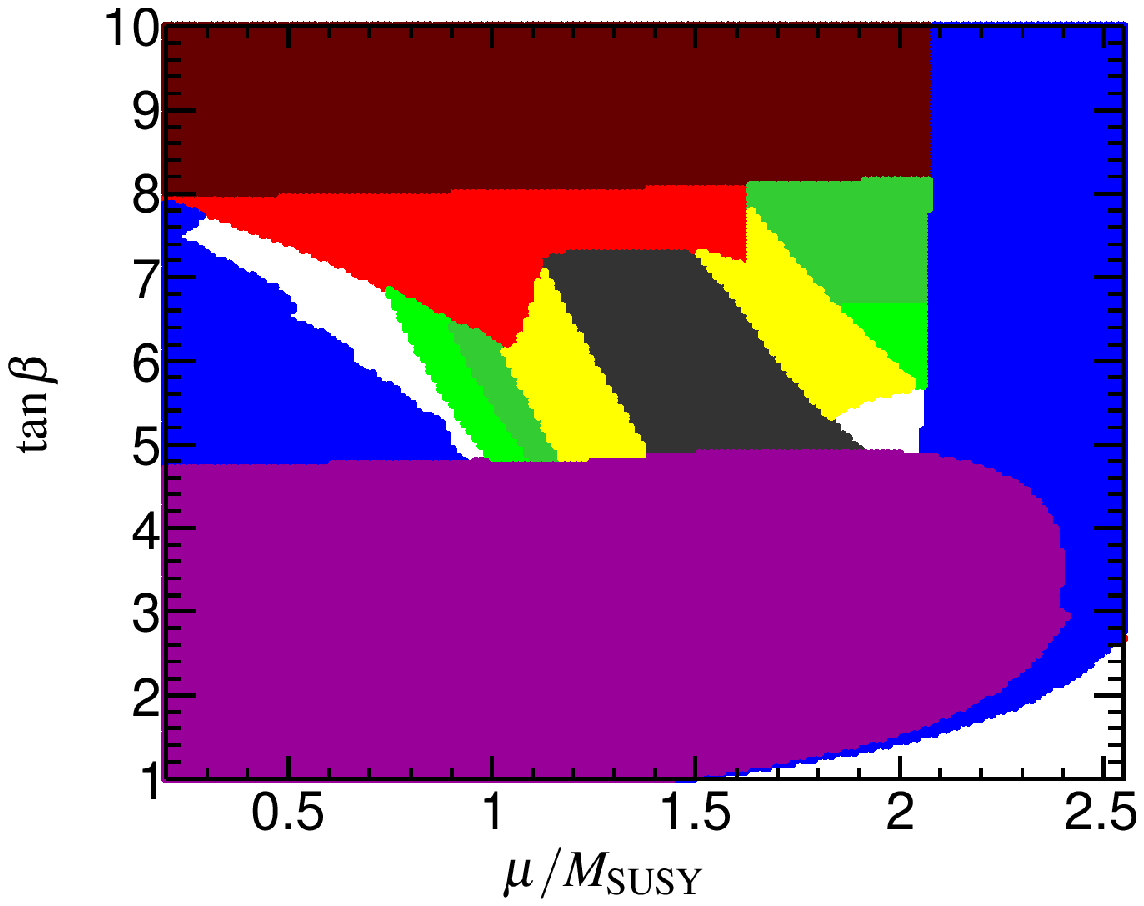}
\includegraphics[width=0.48\columnwidth]{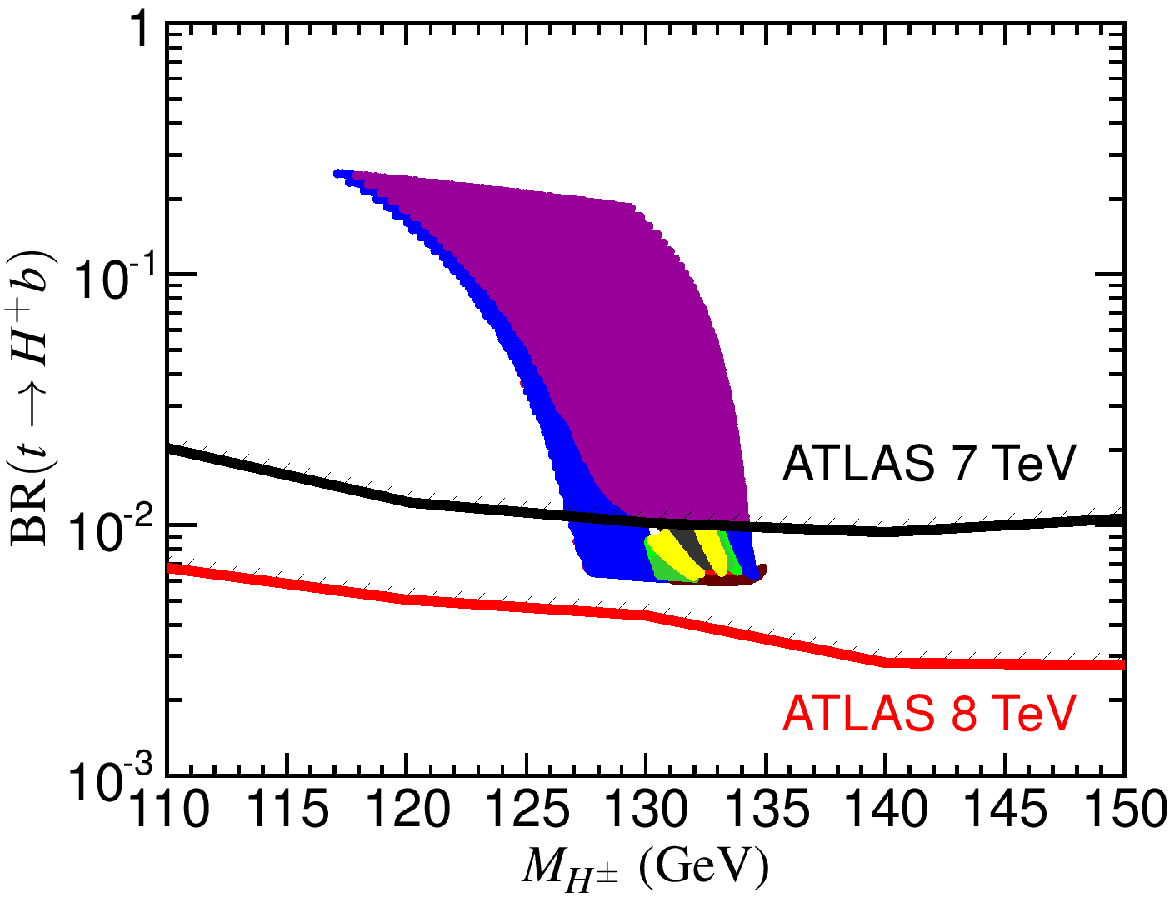}
\caption{Left: exclusion limits from different direct Higgs searches in the MSSM \lowmh\ scenario\protect\cite{Carena:2013qia}. The experimentally allowed regions are shown in green ($M_H=125.5\pm 3\GeV$) and yellow ($M_H=125.5\pm 3\GeV$ and SM-like rates). Right: predictions for $\mathrm{BR}(t\to bH^+)\times \mathrm{BR}(H^+\to \tau^+\nu_\tau)$ in this scenario, with the 7 TeV ATLAS upper limit\protect\cite{Aad:2012tj} (black) and updated 8 TeV results\protect\cite{ATLAS-CONF-2013-090} (red).}
\label{fig:lowMH}
\end{figure}

Another, quite unusual, scenario which can arise is that the observed state is not the lightest state of an extended Higgs sector. An interpretation of the LHC data in terms of the second, heavier, \cp-even Higgs has been proposed in the MSSM\cite{Heinemeyer:2011aa}, as well as in the next-to-minimal model (NMSSM)\cite{Gunion:2012gc,*Belanger:2012tt}. Perhaps surprisingly, the main constraints on these scenarios do typically \emph{not} come from the presence of a lighter Higgs state (with a mass that is often below the SM LEP limit $M_h\gtrsim 114\GeV$), but rather from the presence of additional doublet-like states around $125\GeV$. In the MSSM these are the usual $A$ and $H^\pm$, where in particular the searches for the charged Higgs in top decays poses unavoidable constraints. This is illustrated in Fig.~\ref{fig:lowMH}, which shows as an example the parameter space of the so-called \lowmh\ scenario\cite{Carena:2013qia} (left), and how the prediction for $\mathrm{BR}(t\to bH^+)$ compares to the latest ATLAS 8 TeV limit\cite{ATLAS-CONF-2013-090} (right). As the figure shows, there is not much room anymore for this interpretation in the MSSM. In non-minimal models, where the mass relations between $H^\pm$ and the other Higgs bosons are relaxed, the existence of Higgs bosons lighter than $125\GeV$ remains a viable possibility.

\subsection{Beyond minimal SUSY}
Besides the prospects to continue MSSM heavy Higgs searches with more data in the ``standard'' channels with high sensitivity---mainly for high $\tan\beta$---it has recently been proposed\cite{Djouadi:2013vqa} to extend searches into a regime at low $\tan\beta$ which is usually not accessible in the MSSM. This idea takes seriously the possibility that the stop mass scale is not only unnatural, but really multi-TeV, something which is certainly compatible with all (non-)observations. In this case, the lightest Higgs mass can be made sufficiently heavy also for $\tan\beta\sim 1$--$2$, where the heavy \cp-even scalar, $H$, has dominant decays into $ZZ$, $W^+W^-$, $hh$ or $t\bar{t}$. 
\begin{figure}[b]
\centering
\includegraphics[width=0.4\columnwidth]{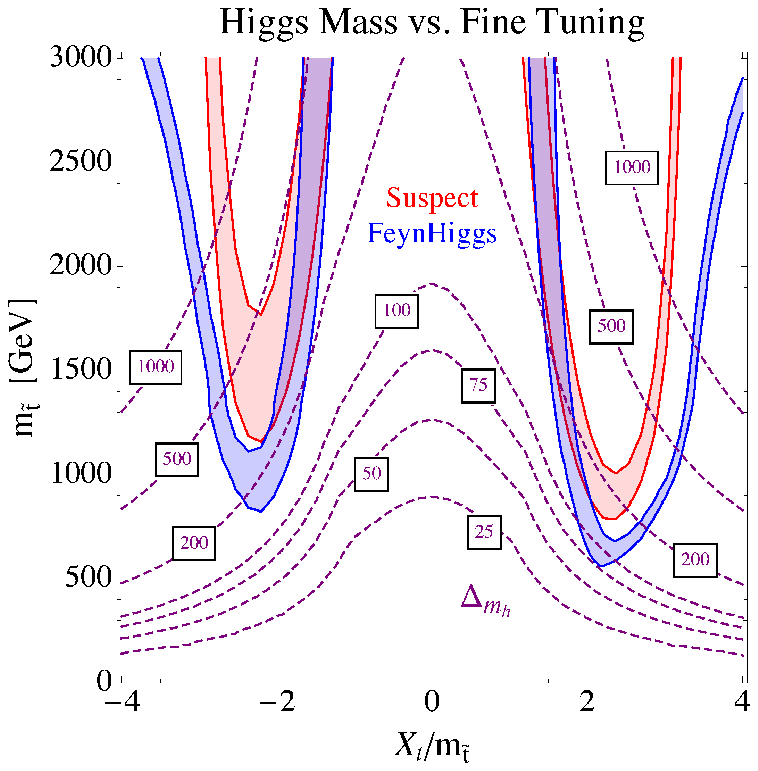}\, \,
\includegraphics[width=0.4\columnwidth]{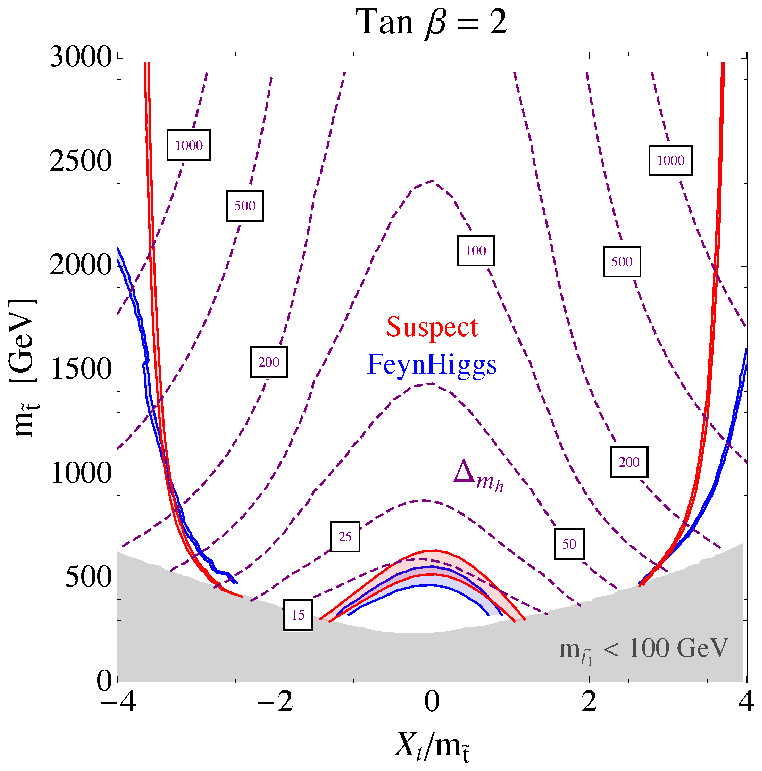}
\caption{Degree of Higgs mass fine-tuning $\Delta m_h$ (contours) from\protect\cite{Hall:2011aa}. Higgs mass predictions (coloured bands) are shown for $m_{\tilde{t}}\equiv M_S$ versus $X_t$ in the MSSM (left), and the NMSSM (right) with $\tan\beta=2$ and $\lambda=0.7$ (the maximal value perturbative up to the GUT scale).}
\label{fig:finetuning}
\end{figure}
While this comes at a high price in the MSSM, it should be noted that the same interesting parameter regime could also be accessible in more natural models with non-minimal supersymmetry, such as the NMSSM. In the NMSSM, a singlet superfield $\hat{S}$ is added to the MSSM, with a superpotential $W^{(3)}_{\mathrm{NMSSM}}=\lambda\hat{S}\hat{H}_u\cdot \hat{H}_d+(\kappa/3)\hat{S}^3$. The resulting spectrum contains seven physical Higgs states and the lightest NMSSM Higgs mass receives an additional tree-level contribution proportional to $\lambda$.
Fig.~\ref{fig:finetuning} shows the degree of fine-tuning resulting from $M_h$, calculated in\cite{Hall:2011aa} for the MSSM (left) and the NMSSM (right). As can be seen from this figure, the Higgs mass fine-tuning can be significantly reduced in the NMSSM. The NMSSM could also offer attractive features from both phenomenological and experimental points-of-view, such as singlet-doublet mixing as an alternative to enhance loop-mediated $h\to\gamma\gamma$ decays\cite{Ellwanger:2010nf,*Cao:2011pg,*Ellwanger:2011aa,*Cao:2012fz,*Benbrik:2012rm} and the existence of light Higgs bosons\cite{Dermisek:2008uu} that can be searched for in different ways\cite{Stal:2011cz,*Ellwanger:2013ova,*Cao:2013gba}. A comprehensive review of phenomenological studies in the NMSSM Higgs sector can be found in\cite{Ellwanger:2011sk}.

\vspace{-1em}
\section{General two-Higgs-doublet models}
An alternative framework to interpret Higgs boson searches, including searches sensitive to new Higgs sectors beyond the SM, is the general two-Higgs-doublet model (2HDM)\cite{Branco:2011iw}. This model in itself does not address the SM hierarchy problem, but it is interesting as a possible low-energy limit of a UV-complete theory (for which the MSSM is one example). The 2HDM offers a large freedom, since there are enough free parameters to specify e.g.~all four physical Higgs masses as input. The current data already leads to non-trivial constraints on the 2HDM parameter space\cite{Ferreira:2013qua,*Belanger:2013xza}. Most analyses so far focus on the 2HDM types I/II with a (softly broken) $Z_2$-symmetry, but there are also studies with more general Yukawa couplings\cite{Celis:2013ixa}. Due to the large freedom in choosing the parameters, the heavy Higgs bosons are less constrained by direct searches than in SUSY models. On the other hand, the absence of additional new states typically leads to strong constraints from flavour physics\cite{Mahmoudi:2009zx}. First experimental LHC results from a 2HDM Higgs search in the $H\to WW$ channel have been presented\cite{ATLAS-CONF-2013-027}, but there are many other channels that could be interesting for a 2HDM interpretation. In case future searches for additional Higgs bosons reveal nothing, this would indicate for the 2HDM---like in the case of the MSSM---that the most probable scenario is the decoupling limit. In this limit, the heavy Higgs states decouple, leaving a single light (SM-like) Higgs boson in the spectrum. To measure the small deviations in the Higgs couplings predicted in this case would likely require the precision of a linear $e^+e^-$ collider\cite{Asner:2013psa}.

Another instance of the 2HDM leading to similar results is the Inert Doublet Model (IDM)\cite{Deshpande:1977rw,*Barbieri:2006dq,*Ma:2006km}. In the IDM, the $Z_2$-symmetry is made exact, rendering the lightest scalar particle stable. In this way the model can accommodate a dark matter (DM) candidate. Recent analyses\cite{Goudelis:2013uca,*Krawczyk:2013jta} demonstrate that this is compatible with the Higgs discovery. However, taking into account data on invisible Higgs decays\cite{ATLASinv}, in most of the allowed parameter space the mass of the DM candidate is $M_{\rm{DM}}>500\GeV$, and it appears difficult to discover any of the additional Higgs bosons at the LHC. The only exception is when the scalar DM exhibits resonant annihilation on the Higgs pole, $M_{\rm{DM}}\gtrsim M_h/2\sim 60$--$80$~GeV, which could be an interesting mass region.
\vspace{-1em}

\section{Fits of the Higgs couplings}
A generic way to test for (absence of) new physics is to parametrize coupling deviations from the SM, and then use the available data to determine constraints on these deviations. Here we focus on deviations in the signal \emph{strengths}, for which a simple ``interim'' framework has been proposed\cite{LHCHiggsCrossSectionWorkingGroup:2012nn,*Heinemeyer:2013tqa}. In the most general case, each SM coupling is assigned a scale factor, $\kappa_i$, defined such that, for example,
\begin{equation}
\sigma(gg\to H)\times \mathrm{BR}(H\to WW) = \frac{\kappa_g^2\kappa^2_W}{\kappa_H^2} \left[\sigma(gg\to H)\times \rm{BR}(H\to WW)\right]_{\mathrm{SM}},
\end{equation}
where $\kappa_H$ is a scale factor for the Higgs total width (which is not measurable at the LHC). This definition allows the $\kappa_i$ to be extracted from data using the most accurate predictions available for the SM production cross sections and decay branching ratios.\footnote{To allow for deviations in the \cp, spin, or tensor structure of the SM Higgs couplings, more elaborate parametrizations are required\cite{Contino:2013kra}.}
Using this framework, different fits to the (SM) measurements can be performed where some of the coupling scale factors are allowed to vary. This strategy has been pursued by several theory groups (there are too many results to list them here), and also directly by the experimental collaborations\cite{ATLAS-CONF-2013-034, CMS-PAS-HIG-13-005}. 

\begin{figure}
\centering
\includegraphics[width=0.6\columnwidth]{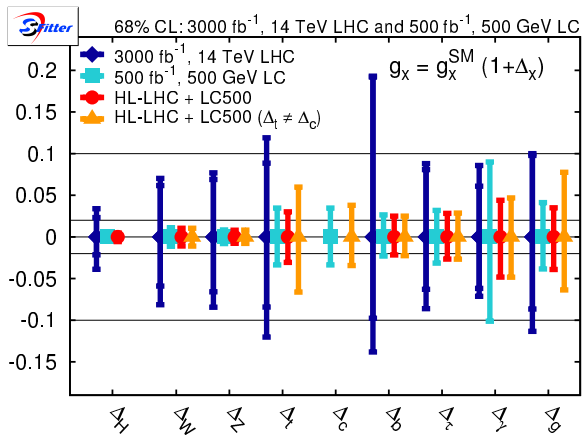}
\caption{Projected sensitivity from\protect\cite{Klute:2013cx} for Higgs coupling measurements (deviations from the SM) at the 14 TeV HL-LHC and a 500 GeV $e^+e^-$ linear collider (LC).}
\label{fig:ILC}
\end{figure}

A significant deviation from the SM point ($\forall \kappa_i: \kappa_i=1$) in one of these coupling fits would signal a shortcoming of the SM to account for the measured data. Currently, there is no such indication, and the constraints on the allowed deviation in individual $\kappa_i$ from highly constrained analyses are of the order $10-20\%$ (but larger if more scale factors are varied). In the absence of direct observations of new physics, it will be a very important goal of the LHC to improve the constraints on the SM Higgs couplings and search for deviations at a greater level of precision. From general arguments, new physics entering at a scale $f$ modifying (some of) the Higgs couplings can be expected to contribute at the level 
\begin{equation}
\kappa_i\simeq 1\pm c\frac{v^2}{f^2},
\end{equation}
where $c$ is a number of $\mathcal{O}(1$--$100\%)$\cite{Baer:2013cma}. This already tells us that percent-level precision is desirable to approach the TeV-scale. While an improvemed precision for the coupling measurements is certainly expected at LHC-300, and even more so at a high-luminosity (HL-LHC) with 3000 fb$^{-1}$, the best precision on the Higgs couplings would only be reached at a lepton collider, such as the ILC\cite{Asner:2013psa,Peskin:2012we}. This statement can be made more quantitative by comparing the expected performance for Higgs coupling determination in different future collider scenarios. In Fig.~\ref{fig:ILC} we show a projection by the SFitter collaboration\cite{Klute:2013cx}. It can be seen from this figure that a precision at the few percent level on nearly all the Higgs couplings can be achieved with a linear collider at $500$~GeV alone, and that further improvement on several couplings, in particular $\Delta_\gamma$, can be reached by combining LHC and ILC results.

\subsection{Composite Higgs models}
\begin{figure}
\centering
\includegraphics[width=0.6\columnwidth]{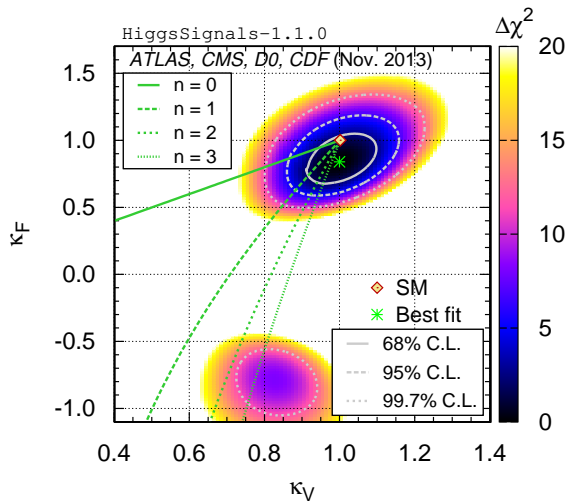}
\vspace{-0.5em}
\caption{Two-parameter fit of Higgs coupling scale factors ($\kappa_V,\kappa_F$) to all available rate measurements performed with \HS\protect\cite{Bechtle:2013xfa}. The colour coding shows $\Delta\chi^2=\chi^2-\chi^2_{\mathrm{min}}$, and the green contours the MCHM predictions for varying compositeness scale $\xi$ and values of $n$.}
\label{fig:mchm}
\end{figure}

An alternative to a weakly coupled natural Higgs sector---as in supersymmetry---would be if the observed scalar is not fundamental, but results as a composite pseudo-Goldstone boson from some broken global symmetry of a new strong interaction\cite{Giudice:2007fh} (much like pions in QCD). These ideas are exemplified by the minimal composite Higgs models\cite{Agashe:2004rs,*Contino:2006qr} (MCHM). In the minimal case, a global SO(5) symmetry is assumed to be broken dynamically down to SO(4) at a scale $f$, leading to a single doublet of $\mathrm{SU}(2)_L$, precisely as in the SM. The compositeness leads to modifications of the SM Higgs couplings, which in the minimal case amounts to universal effects on the coupling scale factors to vector bosons, $\kappa_V$, and fermions, $\kappa_F$. In this scenario, we have explicitly\cite{Pomarol:2012qf}
\begin{equation*}
\kappa_V=\sqrt{1-\xi^2}, \quad
\displaystyle \kappa_F=\frac{1-(1+n)\xi^2}{\sqrt{1-\xi^2}},
\end{equation*}
where $n=0,1,\ldots$ is an integer that depends on the realization and $\xi = v/f$. The scale $f$ plays a role similar to $M_S$ is SUSY theories when it comes to naturalness. A higher symmetry-breaking scale $f$ implies decoupling to the SM limit, but also that the compositeness has no role in a natural solution of the SM hierarchy problem. Conversely, for $f\simeq v$, the SM ``Higgs'' would be a fully composite state of the strong interaction (technicolor limit), an option which is already strongly disfavoured by the data. Current measurements constrain the MCHM, as can be seen in Fig.~\ref{fig:mchm} (for an earlier analysis of this type, see\cite{Falkowski:2013dza}). To produce these constraints we have used the public code \HS\cite{Bechtle:2013xfa}, which provides a generic test for compatibility between Higgs measurements and the predictions of an arbitrary model. Using all available Higgs channels from Tevatron and the LHC, the combined constraints corresponds to a $95\%$ CL~lower limit of $f\gtrsim 500$--$800\GeV$ (depending on $n$), which starts to challenge the naturalness of the theory. 
As already mentioned, to probe deviations in the Higgs couplings up to much higher scales (TeV and beyond) is likely require the full precision of the (upgraded) LHC and/or a lepton collider. However, one particular aspect of the composite Higgs theories is that they quite generically predict the existence of exotic particles around the TeV-scale\cite{Pomarol:2012qf,Marzocca:2012zn}. To discover (or rule out) the existence of these states at the LHC will therefore provide another important test of the Higgs compositeness idea. 

\section{Conclusions}
The new particle discovered by ATLAS and CMS has properties compatible with the SM Higgs boson; no significant deviations from the SM are observed in the data. This has far-reaching implications also for scenarios beyond the SM, which must now face the additional constraint that there should be a SM-like Higgs boson in the spectrum. At the same time, many of these theories (like, for example, supersymmetry) face quite stringent constraints from the non-observation of new particles below the TeV-scale.
Looking at these constraints from a bottom-up perspective on the prospects for extended Higgs scenarios, the picture is however quite different. For theories with a decoupling limit it is fully consistent to expect the lightest state to be a SM-like Higgs boson, and for this particle to become more SM-like the more additional (heavier) degrees of freedom decouple. This still motivates searches for additional Higgs states in e.g.~the MSSM (where the decoupling limit is favoured by global analyses), as well as in the context of general 2HDM. Large effects on the individual rates of the $125$~GeV state are not excluded, with modifications of up to $20$--$100\%$ still allowed in some cases. To explicitly highlight these options and spur experimental activity, updated benchmark scenarios for MSSM Higgs searches have been presented, and new signatures are discussed for heavy Higgs searches. 

Beyond the MSSM, extended supersymmetric scenarios offer a more natural explanation for the ``high'' Higgs mass. The phenomenology of these scenarios can differ substantially from that in the MSSM, and most limits obtained from searches for heavy MSSM Higgs bosons are not directly applicable in this context. The prospects to constrain these models in the near future therefore depend to a large degree on the feasibility to include additional decay modes into the LHC heavy Higgs searches. The situation is similar in the general 2HDM, which offers a parametrization with significant freedom remaining to choose the Higgs masses and couplings.

Precision measurements pertaining to the $125$~GeV state, first at the LHC and later at a future lepton collider, will eventually be sensitive to deviations in individual Higgs rates associated with scales that are not accessible to direct Higgs searches. This could offer the intriguing, but also somewhat frustrating, scenario that the SM is rejected without a direct answer to the question what lies behind. The best prospects for the LHC to prove that the Higgs boson is non-standard is through new discoveries, which would give immediate proof for the existence of physics beyond the SM. Continued Higgs searches in motivated SM extensions should therefore remain a top priority at the LHC.

\section*{Acknowledgments}
I would like to thank the organizers of PIC2013 for the invitation and their kind hospitality. I am also grateful to my collaborators on the various projects from which material has been borrowed for this review; in particular I would like to thank Tim Stefaniak for useful assistance with \HS\ in preparing Fig.~\ref{fig:mchm}.

\bibliographystyle{JHEP}
\bibliography{bsmhiggs}

\ifx\mcitethebibliography\mciteundefinedmacro
\PackageError{JHEP.bst}{mciteplus.sty has not been loaded}
{This bibstyle requires the use of the mciteplus package.}\fi
\providecommand{\href}[2]{#2}\begingroup\raggedright\begin{mcitethebibliography}{10}

\bibitem{ATLASDiscovery}
{ATLAS Collaboration}, G.~Aad {\em et.~al.} {\em Phys.~Lett.~B} {\bf 716}
  (2012) 1, [\href{http://xxx.lanl.gov/abs/1207.7214}{{\tt
  arXiv:1207.7214}}]\relax
\mciteBstWouldAddEndPuncttrue
\mciteSetBstMidEndSepPunct{\mcitedefaultmidpunct}
{\mcitedefaultendpunct}{\mcitedefaultseppunct}\relax
\EndOfBibitem
\bibitem{CMSDiscovery}
{CMS Collaboration}, S.~Chatrchyan {\em et.~al.} {\em Phys.~Lett.~B} {\bf 716}
  (2012) 30, [\href{http://xxx.lanl.gov/abs/1207.7235}{{\tt
  arXiv:1207.7235}}]\relax
\mciteBstWouldAddEndPuncttrue
\mciteSetBstMidEndSepPunct{\mcitedefaultmidpunct}
{\mcitedefaultendpunct}{\mcitedefaultseppunct}\relax
\EndOfBibitem
\bibitem{ATLAS-CONF-2013-034}
{ATLAS Collaboration}. ATLAS-CONF-2013-034\relax
\mciteBstWouldAddEndPuncttrue
\mciteSetBstMidEndSepPunct{\mcitedefaultmidpunct}
{\mcitedefaultendpunct}{\mcitedefaultseppunct}\relax
\EndOfBibitem
\bibitem{CMS-PAS-HIG-13-005}
{CMS Collaboration}. CMS-PAS-HIG-13-005\relax
\mciteBstWouldAddEndPuncttrue
\mciteSetBstMidEndSepPunct{\mcitedefaultmidpunct}
{\mcitedefaultendpunct}{\mcitedefaultseppunct}\relax
\EndOfBibitem
\bibitem{ATLAS-CONF-2013-014}
{ATLAS Collaboration}. ATLAS-CONF-2013-014\relax
\mciteBstWouldAddEndPuncttrue
\mciteSetBstMidEndSepPunct{\mcitedefaultmidpunct}
{\mcitedefaultendpunct}{\mcitedefaultseppunct}\relax
\EndOfBibitem
\bibitem{Degrassi:2012ry}
G.~Degrassi {\em et.~al.} {\em JHEP} {\bf 1208} (2012) 098,
  [\href{http://xxx.lanl.gov/abs/1205.6497}{{\tt arXiv:1205.6497}}]\relax
\mciteBstWouldAddEndPuncttrue
\mciteSetBstMidEndSepPunct{\mcitedefaultmidpunct}
{\mcitedefaultendpunct}{\mcitedefaultseppunct}\relax
\EndOfBibitem
\bibitem{ATLAS-CONF-2013-040}
{ATLAS Collaboration}. ATLAS-CONF-2013-040\relax
\mciteBstWouldAddEndPuncttrue
\mciteSetBstMidEndSepPunct{\mcitedefaultmidpunct}
{\mcitedefaultendpunct}{\mcitedefaultseppunct}\relax
\EndOfBibitem
\bibitem{Chatrchyan:2012jja}
{CMS Collaboration}, S.~Chatrchyan {\em et.~al.} {\em Phys.~Rev.~Lett.} {\bf
  110} (2013) 081803, [\href{http://xxx.lanl.gov/abs/1212.6639}{{\tt
  arXiv:1212.6639}}]\relax
\mciteBstWouldAddEndPuncttrue
\mciteSetBstMidEndSepPunct{\mcitedefaultmidpunct}
{\mcitedefaultendpunct}{\mcitedefaultseppunct}\relax
\EndOfBibitem
\bibitem{ATLASSusyWiki}
 https://twiki.cern.ch/twiki/bin/view/AtlasPublic/SupersymmetryPublicResults;\\
  https://twiki.cern.ch/twiki/bin/view/CMSPublic/PhysicsResultsSUS\relax
\mciteBstWouldAddEndPuncttrue
\mciteSetBstMidEndSepPunct{\mcitedefaultmidpunct}
{\mcitedefaultendpunct}{\mcitedefaultseppunct}\relax
\EndOfBibitem
\bibitem{Brust:2011tb}
C.~Brust, A.~Katz, S.~Lawrence, and R.~Sundrum {\em JHEP} {\bf 1203} (2012)
  103, [\href{http://xxx.lanl.gov/abs/1110.6670}{{\tt arXiv:1110.6670}}]\relax
\mciteBstWouldAddEndPuncttrue
\mciteSetBstMidEndSepPunct{\mcitedefaultmidpunct}
{\mcitedefaultendpunct}{\mcitedefaultseppunct}\relax
\EndOfBibitem
\bibitem{Papucci:2011wy}
M.~Papucci, J.~T. Ruderman, and A.~Weiler {\em JHEP} {\bf 1209} (2012) 035,
  [\href{http://xxx.lanl.gov/abs/1110.6926}{{\tt arXiv:1110.6926}}]\relax
\mciteBstWouldAddEndPuncttrue
\mciteSetBstMidEndSepPunct{\mcitedefaultmidpunct}
{\mcitedefaultendpunct}{\mcitedefaultseppunct}\relax
\EndOfBibitem
\bibitem{Hall:2011aa}
L.~J. Hall, D.~Pinner, and J.~T. Ruderman {\em JHEP} {\bf 1204} (2012) 131,
  [\href{http://xxx.lanl.gov/abs/1112.2703}{{\tt arXiv:1112.2703}}]\relax
\mciteBstWouldAddEndPuncttrue
\mciteSetBstMidEndSepPunct{\mcitedefaultmidpunct}
{\mcitedefaultendpunct}{\mcitedefaultseppunct}\relax
\EndOfBibitem
\bibitem{Heinemeyer:2011aa}
S.~Heinemeyer, O.~St{\aa}l, and G.~Weiglein {\em Phys.~Lett.~B} {\bf 710}
  (2012) 201, [\href{http://xxx.lanl.gov/abs/1112.3026}{{\tt
  arXiv:1112.3026}}]\relax
\mciteBstWouldAddEndPuncttrue
\mciteSetBstMidEndSepPunct{\mcitedefaultmidpunct}
{\mcitedefaultendpunct}{\mcitedefaultseppunct}\relax
\EndOfBibitem
\bibitem{Baer:2011ab}
H.~Baer, V.~Barger, and A.~Mustafayev {\em Phys.~Rev.~D} {\bf 85} (2012)
  075010, [\href{http://xxx.lanl.gov/abs/1112.3017}{{\tt
  arXiv:1112.3017}}]\relax
\mciteBstWouldAddEndPuncttrue
\mciteSetBstMidEndSepPunct{\mcitedefaultmidpunct}
{\mcitedefaultendpunct}{\mcitedefaultseppunct}\relax
\EndOfBibitem
\bibitem{Arbey:2011ab}
A.~Arbey, M.~Battaglia, A.~Djouadi, F.~Mahmoudi, and J.~Quevillon {\em
  Phys.~Lett.~B} {\bf 708} (2012) 162,
  [\href{http://xxx.lanl.gov/abs/1112.3028}{{\tt arXiv:1112.3028}}]\relax
\mciteBstWouldAddEndPuncttrue
\mciteSetBstMidEndSepPunct{\mcitedefaultmidpunct}
{\mcitedefaultendpunct}{\mcitedefaultseppunct}\relax
\EndOfBibitem
\bibitem{Draper:2011aa}
P.~Draper, P.~Meade, M.~Reece, and D.~Shih {\em Phys.~Rev.~D} {\bf 85} (2012)
  095007, [\href{http://xxx.lanl.gov/abs/1112.3068}{{\tt
  arXiv:1112.3068}}]\relax
\mciteBstWouldAddEndPuncttrue
\mciteSetBstMidEndSepPunct{\mcitedefaultmidpunct}
{\mcitedefaultendpunct}{\mcitedefaultseppunct}\relax
\EndOfBibitem
\bibitem{Brummer:2012ns}
F.~Br{\"u}mmer, S.~Kraml, and S.~Kulkarni {\em JHEP} {\bf 1208} (2012) 089,
  [\href{http://xxx.lanl.gov/abs/1204.5977}{{\tt arXiv:1204.5977}}]\relax
\mciteBstWouldAddEndPuncttrue
\mciteSetBstMidEndSepPunct{\mcitedefaultmidpunct}
{\mcitedefaultendpunct}{\mcitedefaultseppunct}\relax
\EndOfBibitem
\bibitem{Carena:1999xa}
M.~S. Carena, S.~Heinemeyer, C.~E.~M. Wagner, and G.~Weiglein
  \href{http://xxx.lanl.gov/abs/hep-ph/9912223}{{\tt hep-ph/9912223}}\relax
\mciteBstWouldAddEndPuncttrue
\mciteSetBstMidEndSepPunct{\mcitedefaultmidpunct}
{\mcitedefaultendpunct}{\mcitedefaultseppunct}\relax
\EndOfBibitem
\bibitem{Schael:2006cr}
{LEP Higgs WG}, S.~Schael {\em et.~al.} {\em Eur.~Phys.~J.~C} {\bf 47} (2006)
  547, [\href{http://xxx.lanl.gov/abs/hep-ex/0602042}{{\tt
  hep-ex/0602042}}]\relax
\mciteBstWouldAddEndPuncttrue
\mciteSetBstMidEndSepPunct{\mcitedefaultmidpunct}
{\mcitedefaultendpunct}{\mcitedefaultseppunct}\relax
\EndOfBibitem
\bibitem{Carena:2013qia}
M.~Carena, S.~Heinemeyer, O.~St{\aa}l, C.~Wagner, and G.~Weiglein {\em
  Eur.~Phys.~J.~C} {\bf 73} (2013) 2552,
  [\href{http://xxx.lanl.gov/abs/1302.7033}{{\tt arXiv:1302.7033}}]\relax
\mciteBstWouldAddEndPuncttrue
\mciteSetBstMidEndSepPunct{\mcitedefaultmidpunct}
{\mcitedefaultendpunct}{\mcitedefaultseppunct}\relax
\EndOfBibitem
\bibitem{Heinemeyer:1998yj}
S.~Heinemeyer, W.~Hollik, and G.~Weiglein {\em Comput.~Phys.~Commun.} {\bf 124}
  (2000) 76, [\href{http://xxx.lanl.gov/abs/hep-ph/9812320}{{\tt
  hep-ph/9812320}}]\relax
\mciteBstWouldAddEndPuncttrue
\mciteSetBstMidEndSepPunct{\mcitedefaultmidpunct}
{\mcitedefaultendpunct}{\mcitedefaultseppunct}\relax
\EndOfBibitem
\bibitem{Heinemeyer:1998np}
 {\em Eur.~Phys.~J.~C} {\bf 9} (1999)
  343, [\href{http://xxx.lanl.gov/abs/hep-ph/9812472}{{\tt
  hep-ph/9812472}}]\relax
\mciteBstWouldAddEndPuncttrue
\mciteSetBstMidEndSepPunct{\mcitedefaultmidpunct}
{\mcitedefaultendpunct}{\mcitedefaultseppunct}\relax
\EndOfBibitem
\bibitem{Degrassi:2002fi}
G.~Degrassi, S.~Heinemeyer, W.~Hollik, P.~Slavich, and G.~Weiglein {\em
  Eur.~Phys.~J.~C} {\bf 28} (2003) 133,
  [\href{http://xxx.lanl.gov/abs/hep-ph/0212020}{{\tt hep-ph/0212020}}]\relax
\mciteBstWouldAddEndPuncttrue
\mciteSetBstMidEndSepPunct{\mcitedefaultmidpunct}
{\mcitedefaultendpunct}{\mcitedefaultseppunct}\relax
\EndOfBibitem
\bibitem{Frank:2006yh}
M.~Frank, T.~Hahn, S.~Heinemeyer, {\em et.~al.} {\em JHEP} {\bf 0702} (2007)
  047, [\href{http://xxx.lanl.gov/abs/hep-ph/0611326}{{\tt
  hep-ph/0611326}}]\relax
\mciteBstWouldAddEndPuncttrue
\mciteSetBstMidEndSepPunct{\mcitedefaultmidpunct}
{\mcitedefaultendpunct}{\mcitedefaultseppunct}\relax
\EndOfBibitem
\bibitem{Bechtle:2008jh}
P.~Bechtle, O.~Brein, S.~Heinemeyer, G.~Weiglein, and K.~E. Williams {\em
  Comput.~Phys.~Commun.} {\bf 181} (2010) 138,
  [\href{http://xxx.lanl.gov/abs/0811.4169}{{\tt arXiv:0811.4169}}]\relax
\mciteBstWouldAddEndPuncttrue
\mciteSetBstMidEndSepPunct{\mcitedefaultmidpunct}
{\mcitedefaultendpunct}{\mcitedefaultseppunct}\relax
\EndOfBibitem
\bibitem{Bechtle:2011sb}
ibid. {\bf 182} (2011) 2605,
  [\href{http://xxx.lanl.gov/abs/1102.1898}{{\tt arXiv:1102.1898}}]\relax
\mciteBstWouldAddEndPuncttrue
\mciteSetBstMidEndSepPunct{\mcitedefaultmidpunct}
{\mcitedefaultendpunct}{\mcitedefaultseppunct}\relax
\EndOfBibitem
\bibitem{Bechtle:2013wla}
P.~Bechtle, O.~Brein, S.~Heinemeyer, O.~St{\aa}l, T.~Stefaniak, and G.~Weiglein
  \href{http://xxx.lanl.gov/abs/1311.0055}{{\tt arXiv:1311.0055}}\relax
\mciteBstWouldAddEndPuncttrue
\mciteSetBstMidEndSepPunct{\mcitedefaultmidpunct}
{\mcitedefaultendpunct}{\mcitedefaultseppunct}\relax
\EndOfBibitem
\bibitem{CMSHig12050}
{CMS Collaboration}. CMS-HIG-12-050\relax
\mciteBstWouldAddEndPuncttrue
\mciteSetBstMidEndSepPunct{\mcitedefaultmidpunct}
{\mcitedefaultendpunct}{\mcitedefaultseppunct}\relax
\EndOfBibitem
\bibitem{Aaij:2013aka}
{LHCb collaboration}, R.~Aaij {\em et.~al.} {\em Phys.~Rev.~Lett.} {\bf 111}
  (2013) 101805, [\href{http://xxx.lanl.gov/abs/1307.5024}{{\tt
  arXiv:1307.5024}}]\relax
\mciteBstWouldAddEndPuncttrue
\mciteSetBstMidEndSepPunct{\mcitedefaultmidpunct}
{\mcitedefaultendpunct}{\mcitedefaultseppunct}\relax
\EndOfBibitem
\bibitem{Cao:2012yn}
J.~Cao, Z.~Heng, J.~M. Yang, and J.~Zhu {\em JHEP} {\bf 1210} (2012) 079,
  [\href{http://xxx.lanl.gov/abs/1207.3698}{{\tt arXiv:1207.3698}}]\relax
\mciteBstWouldAddEndPuncttrue
\mciteSetBstMidEndSepPunct{\mcitedefaultmidpunct}
{\mcitedefaultendpunct}{\mcitedefaultseppunct}\relax
\EndOfBibitem
\bibitem{Bechtle:2012jw}
P.~Bechtle, S.~Heinemeyer, O.~St{\aa}l, T.~Stefaniak, G.~Weiglein, and L.~Zeune
  {\em Eur.~Phys.~J.~C} {\bf 73} (2013) 2354,
  [\href{http://xxx.lanl.gov/abs/1211.1955}{{\tt arXiv:1211.1955}}]\relax
\mciteBstWouldAddEndPuncttrue
\mciteSetBstMidEndSepPunct{\mcitedefaultmidpunct}
{\mcitedefaultendpunct}{\mcitedefaultseppunct}\relax
\EndOfBibitem
\bibitem{Arbey:2012bp}
A.~Arbey, M.~Battaglia, A.~Djouadi, and F.~Mahmoudi {\em Phys.~Lett.~B} {\bf
  720} (2013) 153, [\href{http://xxx.lanl.gov/abs/1211.4004}{{\tt
  arXiv:1211.4004}}]\relax
\mciteBstWouldAddEndPuncttrue
\mciteSetBstMidEndSepPunct{\mcitedefaultmidpunct}
{\mcitedefaultendpunct}{\mcitedefaultseppunct}\relax
\EndOfBibitem
\bibitem{Djouadi:2013lra}
A.~Djouadi \href{http://xxx.lanl.gov/abs/1311.0720}{{\tt
  arXiv:1311.0720}}\relax
\mciteBstWouldAddEndPuncttrue
\mciteSetBstMidEndSepPunct{\mcitedefaultmidpunct}
{\mcitedefaultendpunct}{\mcitedefaultseppunct}\relax
\EndOfBibitem
\bibitem{Carena:2011aa}
M.~Carena, S.~Gori, N.~R. Shah, and C.~E. Wagner {\em JHEP} {\bf 1203} (2012)
  014, [\href{http://xxx.lanl.gov/abs/1112.3336}{{\tt arXiv:1112.3336}}]\relax
\mciteBstWouldAddEndPuncttrue
\mciteSetBstMidEndSepPunct{\mcitedefaultmidpunct}
{\mcitedefaultendpunct}{\mcitedefaultseppunct}\relax
\EndOfBibitem
\bibitem{Carena:2012gp}
M.~Carena, S.~Gori, N.~R. Shah, C.~E. Wagner, and L.-T. Wang {\em JHEP} {\bf
  1207} (2012) 175, [\href{http://xxx.lanl.gov/abs/1205.5842}{{\tt
  arXiv:1205.5842}}]\relax
\mciteBstWouldAddEndPuncttrue
\mciteSetBstMidEndSepPunct{\mcitedefaultmidpunct}
{\mcitedefaultendpunct}{\mcitedefaultseppunct}\relax
\EndOfBibitem
\bibitem{Gunion:2012gc}
J.~F. Gunion, Y.~Jiang, and S.~Kraml {\em Phys.~Rev.~D} {\bf 86} (2012) 071702,
  [\href{http://xxx.lanl.gov/abs/1207.1545}{{\tt arXiv:1207.1545}}]\relax
\mciteBstWouldAddEndPuncttrue
\mciteSetBstMidEndSepPunct{\mcitedefaultmidpunct}
{\mcitedefaultendpunct}{\mcitedefaultseppunct}\relax
\EndOfBibitem
\bibitem{Belanger:2012tt}
G.~Belanger, U.~Ellwanger, J.~F. Gunion, Y.~Jiang, S.~Kraml, {\em et.~al.} {\em
  JHEP} {\bf 1301} (2013) 069, [\href{http://xxx.lanl.gov/abs/1210.1976}{{\tt
  arXiv:1210.1976}}]\relax
\mciteBstWouldAddEndPuncttrue
\mciteSetBstMidEndSepPunct{\mcitedefaultmidpunct}
{\mcitedefaultendpunct}{\mcitedefaultseppunct}\relax
\EndOfBibitem
\bibitem{ATLAS-CONF-2013-090}
{ATLAS Collaboration}. ATLAS-CONF-2013-090\relax
\mciteBstWouldAddEndPuncttrue
\mciteSetBstMidEndSepPunct{\mcitedefaultmidpunct}
{\mcitedefaultendpunct}{\mcitedefaultseppunct}\relax
\EndOfBibitem
\bibitem{Aad:2012tj}
{ATLAS Collaboration}, G.~Aad {\em et.~al.} {\em JHEP} {\bf 1206} (2012) 039,
  [\href{http://xxx.lanl.gov/abs/1204.2760}{{\tt arXiv:1204.2760}}]\relax
\mciteBstWouldAddEndPuncttrue
\mciteSetBstMidEndSepPunct{\mcitedefaultmidpunct}
{\mcitedefaultendpunct}{\mcitedefaultseppunct}\relax
\EndOfBibitem
\bibitem{Djouadi:2013vqa}
A.~Djouadi and J.~Quevillon \href{http://xxx.lanl.gov/abs/1304.1787}{{\tt
  arXiv:1304.1787}}\relax
\mciteBstWouldAddEndPuncttrue
\mciteSetBstMidEndSepPunct{\mcitedefaultmidpunct}
{\mcitedefaultendpunct}{\mcitedefaultseppunct}\relax
\EndOfBibitem
\bibitem{Ellwanger:2010nf}
U.~Ellwanger {\em Phys.~Lett.~B} {\bf 698} (2011) 293,
  [\href{http://xxx.lanl.gov/abs/1012.1201}{{\tt arXiv:1012.1201}}]\relax
\mciteBstWouldAddEndPuncttrue
\mciteSetBstMidEndSepPunct{\mcitedefaultmidpunct}
{\mcitedefaultendpunct}{\mcitedefaultseppunct}\relax
\EndOfBibitem
\bibitem{Cao:2011pg}
J.~Cao, Z.~Heng, T.~Liu, and J.~M. Yang {\em Phys.~Lett.~B} {\bf 703} (2011)
  462, [\href{http://xxx.lanl.gov/abs/1103.0631}{{\tt arXiv:1103.0631}}]\relax
\mciteBstWouldAddEndPuncttrue
\mciteSetBstMidEndSepPunct{\mcitedefaultmidpunct}
{\mcitedefaultendpunct}{\mcitedefaultseppunct}\relax
\EndOfBibitem
\bibitem{Ellwanger:2011aa}
U.~Ellwanger {\em JHEP} {\bf 1203} (2012) 044,
  [\href{http://xxx.lanl.gov/abs/1112.3548}{{\tt arXiv:1112.3548}}]\relax
\mciteBstWouldAddEndPuncttrue
\mciteSetBstMidEndSepPunct{\mcitedefaultmidpunct}
{\mcitedefaultendpunct}{\mcitedefaultseppunct}\relax
\EndOfBibitem
\bibitem{Cao:2012fz}
J.-J. Cao, Z.-X. Heng, J.~M. Yang, Y.-M. Zhang, and J.-Y. Zhu {\em JHEP} {\bf
  1203} (2012) 086, [\href{http://xxx.lanl.gov/abs/1202.5821}{{\tt
  arXiv:1202.5821}}]\relax
\mciteBstWouldAddEndPuncttrue
\mciteSetBstMidEndSepPunct{\mcitedefaultmidpunct}
{\mcitedefaultendpunct}{\mcitedefaultseppunct}\relax
\EndOfBibitem
\bibitem{Benbrik:2012rm}
R.~Benbrik, M.~Gomez~Bock, S.~Heinemeyer, {\em et.~al.} {\em Eur.~Phys.~J.~C}
  {\bf 72} (2012) [\href{http://xxx.lanl.gov/abs/1207.1096}{{\tt
  arXiv:1207.1096}}]\relax
\mciteBstWouldAddEndPuncttrue
\mciteSetBstMidEndSepPunct{\mcitedefaultmidpunct}
{\mcitedefaultendpunct}{\mcitedefaultseppunct}\relax
\EndOfBibitem
\bibitem{Dermisek:2008uu}
R.~Dermisek and J.~F. Gunion {\em Phys.~Rev.~D} {\bf 79} (2009) 055014,
  [\href{http://xxx.lanl.gov/abs/0811.3537}{{\tt arXiv:0811.3537}}]\relax
\mciteBstWouldAddEndPuncttrue
\mciteSetBstMidEndSepPunct{\mcitedefaultmidpunct}
{\mcitedefaultendpunct}{\mcitedefaultseppunct}\relax
\EndOfBibitem
\bibitem{Stal:2011cz}
O.~St{\aa}l and G.~Weiglein {\em JHEP} {\bf 1201} (2012) 071,
  [\href{http://xxx.lanl.gov/abs/1108.0595}{{\tt arXiv:1108.0595}}]\relax
\mciteBstWouldAddEndPuncttrue
\mciteSetBstMidEndSepPunct{\mcitedefaultmidpunct}
{\mcitedefaultendpunct}{\mcitedefaultseppunct}\relax
\EndOfBibitem
\bibitem{Ellwanger:2013ova}
U.~Ellwanger {\em JHEP} {\bf 1308} (2013) 077,
  [\href{http://xxx.lanl.gov/abs/1306.5541}{{\tt arXiv:1306.5541}}]\relax
\mciteBstWouldAddEndPuncttrue
\mciteSetBstMidEndSepPunct{\mcitedefaultmidpunct}
{\mcitedefaultendpunct}{\mcitedefaultseppunct}\relax
\EndOfBibitem
\bibitem{Cao:2013gba}
J.~Cao, F.~Ding, C.~Han, J.~M. Yang, and J.~Zhu {\em JHEP} {\bf 1311} (2013)
  018, [\href{http://xxx.lanl.gov/abs/1309.4939}{{\tt arXiv:1309.4939}}]\relax
\mciteBstWouldAddEndPuncttrue
\mciteSetBstMidEndSepPunct{\mcitedefaultmidpunct}
{\mcitedefaultendpunct}{\mcitedefaultseppunct}\relax
\EndOfBibitem
\bibitem{Ellwanger:2011sk}
U.~Ellwanger {\em Eur.~Phys.~J.~C} {\bf 71} (2011) 1782,
  [\href{http://xxx.lanl.gov/abs/1108.0157}{{\tt arXiv:1108.0157}}]\relax
\mciteBstWouldAddEndPuncttrue
\mciteSetBstMidEndSepPunct{\mcitedefaultmidpunct}
{\mcitedefaultendpunct}{\mcitedefaultseppunct}\relax
\EndOfBibitem
\bibitem{Branco:2011iw}
G.~Branco, P.~Ferreira, Lavoura, {\em et.~al.} {\em Phys.~Rept.} {\bf 516}
  (2012) 1, [\href{http://xxx.lanl.gov/abs/1106.0034}{{\tt
  arXiv:1106.0034}}]\relax
\mciteBstWouldAddEndPuncttrue
\mciteSetBstMidEndSepPunct{\mcitedefaultmidpunct}
{\mcitedefaultendpunct}{\mcitedefaultseppunct}\relax
\EndOfBibitem
\bibitem{Ferreira:2013qua}
P.~Ferreira, R.~Santos, M.~Sher, and J.~P. Silva
  \href{http://xxx.lanl.gov/abs/1305.4587}{{\tt arXiv:1305.4587}}\relax
\mciteBstWouldAddEndPuncttrue
\mciteSetBstMidEndSepPunct{\mcitedefaultmidpunct}
{\mcitedefaultendpunct}{\mcitedefaultseppunct}\relax
\EndOfBibitem
\bibitem{Belanger:2013xza}
G.~Belanger, B.~Dumont, U.~Ellwanger, J.~Gunion, and S.~Kraml {\em
  Phys.~Rev.~D} {\bf 88} (2013) 075008,
  [\href{http://xxx.lanl.gov/abs/1306.2941}{{\tt arXiv:1306.2941}}]\relax
\mciteBstWouldAddEndPuncttrue
\mciteSetBstMidEndSepPunct{\mcitedefaultmidpunct}
{\mcitedefaultendpunct}{\mcitedefaultseppunct}\relax
\EndOfBibitem
\bibitem{Celis:2013ixa}
A.~Celis, V.~Ilisie, and A.~Pich \href{http://xxx.lanl.gov/abs/1310.7941}{{\tt
  arXiv:1310.7941}}\relax
\mciteBstWouldAddEndPuncttrue
\mciteSetBstMidEndSepPunct{\mcitedefaultmidpunct}
{\mcitedefaultendpunct}{\mcitedefaultseppunct}\relax
\EndOfBibitem
\bibitem{Mahmoudi:2009zx}
F.~Mahmoudi and O.~St{\aa}l {\em Phys.~Rev.~D} {\bf 81} (2010) 035016,
  [\href{http://xxx.lanl.gov/abs/0907.1791}{{\tt arXiv:0907.1791}}]\relax
\mciteBstWouldAddEndPuncttrue
\mciteSetBstMidEndSepPunct{\mcitedefaultmidpunct}
{\mcitedefaultendpunct}{\mcitedefaultseppunct}\relax
\EndOfBibitem
\bibitem{ATLAS-CONF-2013-027}
{ATLAS Collaboration}. ATLAS-CONF-2013-027\relax
\mciteBstWouldAddEndPuncttrue
\mciteSetBstMidEndSepPunct{\mcitedefaultmidpunct}
{\mcitedefaultendpunct}{\mcitedefaultseppunct}\relax
\EndOfBibitem
\bibitem{Asner:2013psa}
D.~Asner, T.~Barklow, C.~Calancha, K.~Fujii, N.~Graf, {\em et.~al.}
  \href{http://xxx.lanl.gov/abs/1310.0763}{{\tt arXiv:1310.0763}}\relax
\mciteBstWouldAddEndPuncttrue
\mciteSetBstMidEndSepPunct{\mcitedefaultmidpunct}
{\mcitedefaultendpunct}{\mcitedefaultseppunct}\relax
\EndOfBibitem
\bibitem{Deshpande:1977rw}
N.~G. Deshpande and E.~Ma {\em Phys.~Rev.~D} {\bf 18} (1978) 2574\relax
\mciteBstWouldAddEndPuncttrue
\mciteSetBstMidEndSepPunct{\mcitedefaultmidpunct}
{\mcitedefaultendpunct}{\mcitedefaultseppunct}\relax
\EndOfBibitem
\bibitem{Barbieri:2006dq}
R.~Barbieri, L.~J. Hall, and V.~S. Rychkov {\em Phys.~Rev.~D} {\bf 74} (2006)
  015007, [\href{http://xxx.lanl.gov/abs/hep-ph/0603188}{{\tt
  hep-ph/0603188}}]\relax
\mciteBstWouldAddEndPuncttrue
\mciteSetBstMidEndSepPunct{\mcitedefaultmidpunct}
{\mcitedefaultendpunct}{\mcitedefaultseppunct}\relax
\EndOfBibitem
\bibitem{Ma:2006km}
E.~Ma {\em Phys.~Rev.~D} {\bf 73} (2006) 077301,
  [\href{http://xxx.lanl.gov/abs/hep-ph/0601225}{{\tt hep-ph/0601225}}]\relax
\mciteBstWouldAddEndPuncttrue
\mciteSetBstMidEndSepPunct{\mcitedefaultmidpunct}
{\mcitedefaultendpunct}{\mcitedefaultseppunct}\relax
\EndOfBibitem
\bibitem{Goudelis:2013uca}
A.~Goudelis, B.~Herrmann, and O.~St{\aa}l {\em JHEP} {\bf 1309} (2013) 106,
  [\href{http://xxx.lanl.gov/abs/1303.3010}{{\tt arXiv:1303.3010}}]\relax
\mciteBstWouldAddEndPuncttrue
\mciteSetBstMidEndSepPunct{\mcitedefaultmidpunct}
{\mcitedefaultendpunct}{\mcitedefaultseppunct}\relax
\EndOfBibitem
\bibitem{Krawczyk:2013jta}
M.~Krawczyk, D.~Sokolowska, P.~Swaczyna, and B.~Swiezewska {\em JHEP} {\bf
  1309} (2013) 055, [\href{http://xxx.lanl.gov/abs/1305.6266}{{\tt
  arXiv:1305.6266}}]\relax
\mciteBstWouldAddEndPuncttrue
\mciteSetBstMidEndSepPunct{\mcitedefaultmidpunct}
{\mcitedefaultendpunct}{\mcitedefaultseppunct}\relax
\EndOfBibitem
\bibitem{ATLASinv}
{ATLAS Collaboration}. ATLAS-CONF-2013-011\relax
\mciteBstWouldAddEndPuncttrue
\mciteSetBstMidEndSepPunct{\mcitedefaultmidpunct}
{\mcitedefaultendpunct}{\mcitedefaultseppunct}\relax
\EndOfBibitem
\bibitem{LHCHiggsCrossSectionWorkingGroup:2012nn}
{LHC Higgs Cross Section Working Group}
  \href{http://xxx.lanl.gov/abs/1209.0040}{{\tt arXiv:1209.0040}}\relax
\mciteBstWouldAddEndPuncttrue
\mciteSetBstMidEndSepPunct{\mcitedefaultmidpunct}
{\mcitedefaultendpunct}{\mcitedefaultseppunct}\relax
\EndOfBibitem
\bibitem{Heinemeyer:2013tqa}
 \href{http://xxx.lanl.gov/abs/1307.1347}{{\tt arXiv:1307.1347}}\relax
\mciteBstWouldAddEndPuncttrue
\mciteSetBstMidEndSepPunct{\mcitedefaultmidpunct}
{\mcitedefaultendpunct}{\mcitedefaultseppunct}\relax
\EndOfBibitem
\bibitem{Contino:2013kra}
R.~Contino, M.~Ghezzi, C.~Grojean, M.~M{\"u}hlleitner, and M.~Spira {\em JHEP}
  {\bf 1307} (2013) 035, [\href{http://xxx.lanl.gov/abs/1303.3876}{{\tt
  arXiv:1303.3876}}]\relax
\mciteBstWouldAddEndPuncttrue
\mciteSetBstMidEndSepPunct{\mcitedefaultmidpunct}
{\mcitedefaultendpunct}{\mcitedefaultseppunct}\relax
\EndOfBibitem
\bibitem{Baer:2013cma}
H.~Baer, T.~Barklow, K.~Fujii, Y.~Gao, A.~Hoang, {\em et.~al.}
  \href{http://xxx.lanl.gov/abs/1306.6352}{{\tt arXiv:1306.6352}}\relax
\mciteBstWouldAddEndPuncttrue
\mciteSetBstMidEndSepPunct{\mcitedefaultmidpunct}
{\mcitedefaultendpunct}{\mcitedefaultseppunct}\relax
\EndOfBibitem
\bibitem{Peskin:2012we}
M.~E. Peskin \href{http://xxx.lanl.gov/abs/1207.2516}{{\tt
  arXiv:1207.2516}}\relax
\mciteBstWouldAddEndPuncttrue
\mciteSetBstMidEndSepPunct{\mcitedefaultmidpunct}
{\mcitedefaultendpunct}{\mcitedefaultseppunct}\relax
\EndOfBibitem
\bibitem{Klute:2013cx}
M.~Klute, R.~Lafaye, T.~Plehn, M.~Rauch, and D.~Zerwas {\em Europhys.~Lett.}
  {\bf 101} (2013) 51001, [\href{http://xxx.lanl.gov/abs/1301.1322}{{\tt
  arXiv:1301.1322}}]\relax
\mciteBstWouldAddEndPuncttrue
\mciteSetBstMidEndSepPunct{\mcitedefaultmidpunct}
{\mcitedefaultendpunct}{\mcitedefaultseppunct}\relax
\EndOfBibitem
\bibitem{Giudice:2007fh}
G.~Giudice, C.~Grojean, A.~Pomarol, and R.~Rattazzi {\em JHEP} {\bf 0706}
  (2007) 045, [\href{http://xxx.lanl.gov/abs/hep-ph/0703164}{{\tt
  hep-ph/0703164}}]\relax
\mciteBstWouldAddEndPuncttrue
\mciteSetBstMidEndSepPunct{\mcitedefaultmidpunct}
{\mcitedefaultendpunct}{\mcitedefaultseppunct}\relax
\EndOfBibitem
\bibitem{Agashe:2004rs}
K.~Agashe, R.~Contino, and A.~Pomarol {\em Nucl.~Phys.~B} {\bf 719} (2005) 165,
  [\href{http://xxx.lanl.gov/abs/hep-ph/0412089}{{\tt hep-ph/0412089}}]\relax
\mciteBstWouldAddEndPuncttrue
\mciteSetBstMidEndSepPunct{\mcitedefaultmidpunct}
{\mcitedefaultendpunct}{\mcitedefaultseppunct}\relax
\EndOfBibitem
\bibitem{Contino:2006qr}
R.~Contino, L.~Da~Rold, and A.~Pomarol {\em Phys.~Rev.~D} {\bf 75} (2007)
  055014, [\href{http://xxx.lanl.gov/abs/hep-ph/0612048}{{\tt
  hep-ph/0612048}}]\relax
\mciteBstWouldAddEndPuncttrue
\mciteSetBstMidEndSepPunct{\mcitedefaultmidpunct}
{\mcitedefaultendpunct}{\mcitedefaultseppunct}\relax
\EndOfBibitem
\bibitem{Pomarol:2012qf}
A.~Pomarol and F.~Riva {\em JHEP} {\bf 1208} (2012) 135,
  [\href{http://xxx.lanl.gov/abs/1205.6434}{{\tt arXiv:1205.6434}}]\relax
\mciteBstWouldAddEndPuncttrue
\mciteSetBstMidEndSepPunct{\mcitedefaultmidpunct}
{\mcitedefaultendpunct}{\mcitedefaultseppunct}\relax
\EndOfBibitem
\bibitem{Falkowski:2013dza}
A.~Falkowski, F.~Riva, and A.~Urbano {\em JHEP} {\bf 1311} (2013) 111,
  [\href{http://xxx.lanl.gov/abs/1303.1812}{{\tt arXiv:1303.1812}}]\relax
\mciteBstWouldAddEndPuncttrue
\mciteSetBstMidEndSepPunct{\mcitedefaultmidpunct}
{\mcitedefaultendpunct}{\mcitedefaultseppunct}\relax
\EndOfBibitem
\bibitem{Bechtle:2013xfa}
P.~Bechtle, S.~Heinemeyer, O.~St{\aa}l, T.~Stefaniak, and G.~Weiglein
  \href{http://xxx.lanl.gov/abs/1305.1933}{{\tt arXiv:1305.1933}}\relax
\mciteBstWouldAddEndPuncttrue
\mciteSetBstMidEndSepPunct{\mcitedefaultmidpunct}
{\mcitedefaultendpunct}{\mcitedefaultseppunct}\relax
\EndOfBibitem
\bibitem{Marzocca:2012zn}
D.~Marzocca, M.~Serone, and J.~Shu {\em JHEP} {\bf 1208} (2012) 013,
  [\href{http://xxx.lanl.gov/abs/1205.0770}{{\tt arXiv:1205.0770}}]\relax
\mciteBstWouldAddEndPuncttrue
\mciteSetBstMidEndSepPunct{\mcitedefaultmidpunct}
{\mcitedefaultendpunct}{\mcitedefaultseppunct}\relax
\EndOfBibitem
\end{mcitethebibliography}\endgroup

\end{document}